# Minimum-Information LQG Control
# Part II: Retentive Controllers

Roy Fox[†] and Naftali Tishby[†]

*Abstract*— Retentive (memory-utilizing) sensing-acting agents may operate under limitations on the communication between their sensing, memory and acting components, requiring them to trade off the external cost that they incur with the capacity of their communication channels. In this paper we formulate this problem as a sequential rate-distortion problem of minimizing the rate of information required for the controller's operation under a constraint on its external cost. We reduce this bounded retentive control problem to the memoryless one, studied in Part I of this work [1], by viewing the memory reader as one more sensor and the memory writer as one more actuator. We further investigate the structure of the resulting optimal solution and demonstrate its interesting phenomenology.

## I. INTRODUCTION

In a feedback-control system, the internal state of the agent interacts with the external state of the world through sensors that pay attention to the agent's environment and actuators that apply intention to it, in a perception-action cycle [2]. This interaction is limited by external constraints on observability and controllability, as well as internal constraints on the information-processing resources available to the controller.

In Part I of this work [1], we focused on memoryless controllers that have no internal memory and can only attend to their most recent input observation. We discussed how the communication from the sensor to the actuator is central to the agent's ability to act upon the perceived information. The degree of this attention, measured by the amount of Shannon information about the input observation that is utilized in the output control, is a lower bound on the required capacity of the communication channel between the controller's sensor and its actuator. When this capacity for internal communication is limited, the agent needs to trade off some external cost for reducing the rate at which it transmits information.

A related but often overlooked resource is memory bandwidth. We can think of memory as a communication channel from the past internal state of the controller to its future internal state. When memory resources are remote, communication constraints apply to them as well. Even local memory is limited by its capacity to store information and by the capacity of the internal communication channels to and from the memory components. This limitation is evidenced by the hierarchical design of memory in modern digital computers, which places larger capacity on the channels to closer but smaller cache memory components [3].

When the controller is retentive (memory-utilizing), it does maintain an internal memory state which can have information on more than the most recent observation. As in Part I, our guiding principle in this work is to measure the information complexity of the controller's internal representation by asking *"How much information does the controller have on the past?"*. The retentive controller receives information of the past through both memory and sensory channels (Figure 2) and the amount of information that it keeps of the past is a lower bound on the total capacity of both these channels [4].

In a sense, we can consider the reader of the memory state to be one more sensor and the writer of the memory state to be one more actuator. This suggests a reduction from the retentive case to the memoryless case, in which the memory state is considered external and part of the world state [5], [6]. This memory component is fully observable, fully controllable, has no process noise and incurs no cost. Rather than redevelop our results for the retentive controllers similarly to Part I, this reduction allows us to reuse those results and underlines the structure of the solution.

In this paper we make two contributions. First, we present a method for the design of controllers that are optimal under a constraint on both their memory and sensory channel capacity. To our knowledge, this is the first explicit treatment of the channel capacity of the memory process in the context of continuous state-space systems.

Second, we provide a reduction from the problem of bounded retentive control to the problem of bounded memoryless control. This reduction is conceptually convenient and constructive, allowing us to treat both problems using the same framework and providing insight into the structure of the optimal retentive controller.

In Section II we define the LQG task and restate the results of Part I. In Section III we present the retentive control model, its reduction to memoryless control and the structure of the resulting optimal solution. In Section IV we illustrate our results with an example.

## II. PRELIMINARIES

### A. Control task

We consider the same closed-loop control problem detailed in Part I [1, Section II]. In time $t$, a plant in state $x_t \in \mathbb{R}^n$ emits an observation $y_t \in \mathbb{R}^k$, takes in a control input $u_t \in \mathbb{R}^\ell$ and undergoes a stochastic state transition. We focus on discrete-time systems with linear dynamics,

[†]School of Computer Science and Engineering, The Hebrew University, {royf,tishby}@cs.huji.ac.il

[*]This work was supported by the DARPA MSEE Program, the Gatsby Charitable Foundation, the Israel Science Foundation and the Intel ICRI-CI Institute

Gaussian noise and quadratic cost rate (LQG). For simplicity, all elements are taken to be homogeneous, i.e. centered at the origin, and time-invariant. We note that all our results hold without these assumptions, with the appropriate adjustments, as usual in LQG problems [7].

*Definition 1:* A linear-Gaussian time-invariant (LTI) plant $\langle A, B, C, \Sigma_\xi, \Sigma_\epsilon \rangle$ has state dynamics

$$x_{t+1} = Ax_t + Bu_t + \xi_t; \qquad \xi_t \sim \mathcal{N}(0, \Sigma_\xi), \quad (1)$$

where $A \in \mathbb{R}^{n \times n}$, $B \in \mathbb{R}^{n \times \ell}$, $0 \preceq \Sigma_\xi \in \mathbb{S}^n_+$ and $\xi_t$ is independent of $(x^t, y^t, u^t)$. The observation dynamics are

$$y_t = Cx_t + \epsilon_t; \qquad \epsilon_t \sim \mathcal{N}(0, \Sigma_\epsilon), \quad (2)$$

where $C \in \mathbb{R}^{k \times n}$, $\Sigma_\epsilon \in \mathbb{S}^k_+$ and $\epsilon_t$ is independent of $(y^{t-1}, u^{t-1}, x^t)$, where we denote $x^t = \{x_\tau\}_{\tau \leq t}$, etc.

*Definition 2:* A linear-quadratic-Gaussian (LQG) task $\langle A, B, C, \Sigma_\xi, \Sigma_\epsilon, Q, R \rangle$ involves a LTI plant and the cost rate

$$\mathcal{J}_t = \tfrac{1}{2}(x_t^\intercal Q x_t + u_t^\intercal R u_t),$$

where $Q \in \mathbb{S}^n_+$ and $R \in \mathbb{S}^\ell_+$. The task is to achieve a low long-term average expected cost rate, with respect to the distribution induced by the plant and the controller $\pi$

$$\mathcal{J}_\pi = \limsup_{T \to \infty} \frac{1}{T} \sum_{t=1}^{T} \mathbb{E}_\pi[\mathcal{J}_t].$$

As motivated in Part I, we are particularly interested in linear-Gaussian time-invariant (LTI) controllers, which induce, jointly with a LTI plant, a stationary Gaussian process, independent of any initial conditions. With $\Sigma_x \in \mathbb{S}^n_+$, $\Sigma_y \in \mathbb{S}^k_+$ and $\Sigma_u \in \mathbb{S}^\ell_+$, respectively the stationary covariances of the state, the observation and the control, we have

$$\Sigma_y = C \Sigma_x C^\intercal + \Sigma_\epsilon,$$

and the reverse relation

$$x_t = Ky_t + \kappa_t; \qquad \kappa_t \sim \mathcal{N}(0, \Sigma_\kappa)$$
$$K = \Sigma_x C^\intercal \Sigma_y^\dagger$$
$$\Sigma_\kappa = \Sigma_x - \Sigma_x C^\intercal \Sigma_y^\dagger C \Sigma_x,$$

with $\cdot^\dagger$ the Moore-Penrose pseudoinverse. Assuming that the process has mean 0, the stationary expected cost rate is

$$\mathcal{J}_\pi = \tfrac{1}{2}(\text{tr}(Q \Sigma_x) + \text{tr}(R \Sigma_u)).$$

*B. Bounded memoryless control*

In this section we restate the main result of Part I [1, Section IV].

*Definition 3:* A memoryless linear-Gaussian time-invariant (LTI) controller has control law of the form

$$u_t = H y_t + \eta_t; \qquad \eta_t \sim \mathcal{N}(0, \Sigma_\eta), \quad (3)$$

where $H \in \mathbb{R}^{\ell \times k}$, $\Sigma_\eta \in \mathbb{S}^\ell_+$ and $\eta_t$ is independent of $(u^{t-1}, x^t, y^t)$.

The controller is bounded and operates under limitations on its capacity to process the observation and produce the control. Namely, with the Shannon information rate

$$\mathcal{I}_t = \mathbb{I}[y_t; u_t] = \mathbb{E}\left[\log \frac{f(y_t, u_t)}{f(y_t) f(u_t)}\right], \quad (4)$$

where $f$ denotes the various probability density functions, as indicated by their arguments, we are interested in a LTI controller $\pi$ that minimizes the long-term average rate

$$\mathcal{I}_\pi = \limsup_{T \to \infty} \frac{1}{T} \sum_{t=1}^{T} \mathcal{I}_t, \quad (5)$$

under the constraint that it achieves some guarantee level $c$ of expected cost rate.

*Problem 1:* Given a LQG task, the bounded memoryless LTI controller optimization problem is

$$\min_\pi \quad \mathcal{I}_\pi$$
$$\text{s.t.} \quad \mathcal{J}_\pi \leq c,$$

with $\mathcal{I}_\pi$ as in (5), where $\mathcal{I}_t = \mathbb{I}[y_t; u_t]$, and with $u_t$ as in (3).

To solve the optimization problem, we consider the minimum mean square error (MMSE) estimators

$$\hat{x}_{y_t} = \mathbb{E}[x_t | y_t] = K y_t$$
$$\hat{x}_{u_t} = \mathbb{E}[x_t | u_t] = \Sigma_{x;u} \Sigma_u^\dagger u_t,$$

respectively for the state given the observation and the control. Since $\hat{x}_{u_t}$ is a sufficient statistic of $u_t$ for $x_t$, we can reverse their causality, basing $u_t$ on $\hat{x}_{u_t}$ instead of vice versa. This puts the control law in the form

$$\hat{x}_{y_t} = K y_t$$
$$\hat{x}_{u_t} = W \hat{x}_{y_t} + \omega_t; \qquad \omega_t \sim \mathcal{N}(0, \Sigma_\omega)$$
$$u_t = L \hat{x}_{u_t}.$$

The optimal memoryless controller satisfies the conditions of Theorem 1 in Part I, Section IV-A, restated below in algorithmic form. To numerically find the optimal solution, we can interpret these conditions as update equations, which we apply iteratively until a fixed point is reached.

We split the equations into three parts, a forward iteration (Algorithm 1) updating the marginal distributions, a backward iteration (Algorithm 2) updating the cost-to-go and the control policy, and an eigenvalue decomposition (EVD) for finding the control-based estimator covariance (Algorithm 3). We can alternate between Algorithms 1, 2 and 3, iterating until the solution converges to a fixed point of the equations.

III. BOUNDED RETENTIVE CONTROLLERS

*A. Control model*

In this section we discuss retentive (memory-utilizing) controllers with bounded communication resources. A retentive controller has an internal memory state $z_t$ in some space $\mathcal{Z}$. The memory allows the controller to output a control that indirectly depends on past input observations rather than only on the most recent observation. The controller takes as input an observation $y_t$ and outputs a control $u_t$, while also making a memory state transition from $z_{t-1}$ to $z_t$. Thus, in each time step, there are two inputs, $z_{t-1}$ and $y_t$, and two outputs, $z_t$ and $u_t$.

*Definition 4:* A controller is retentive if it satisfies the following independence properties:

**Algorithm 1** Forward iteration
**function** FORWARD($\Sigma_x, \Sigma_{\hat{x}_u}, L$)
 Update
$$\Sigma_x \leftarrow (A+BL)\Sigma_{\hat{x}_u}(A+BL)^\intercal$$
$$\qquad + A(\Sigma_x - \Sigma_{\hat{x}_u})A^\intercal + \Sigma_\xi$$
$$\Sigma_y \leftarrow C\Sigma_x C^\intercal + \Sigma_\epsilon$$
$$K \leftarrow \Sigma_x C^\intercal \Sigma_y^\dagger$$
$$\Sigma_{\hat{x}_y} \leftarrow K\Sigma_y K^\intercal$$
**end function**

**Algorithm 2** Backward iteration
**function** BACKWARD($\Sigma_{\hat{x}_y}, \Sigma_{\hat{x}_u}, K, S; \beta$)
 Update
$$M \leftarrow \beta^{-1} C^\intercal K^\intercal (\Sigma_{\hat{x}_y|\hat{x}_u}^\dagger - \Sigma_{\hat{x}_y}^\dagger)KC$$
$$S \leftarrow Q + A^\intercal SA - M$$
$$L \leftarrow -(R + B^\intercal SB)^\dagger B^\intercal SA$$
$$N \leftarrow L^\intercal (R + B^\intercal SB)L$$
**end function**

**Algorithm 3** Activation of control-based estimator modes
**function** ACTIVATION($\Sigma_{\hat{x}_y}, N; \beta$)
 Update
$$V, \Lambda \leftarrow \text{EVD}(\Sigma_{\hat{x}_y}^{1/2} N \Sigma_{\hat{x}_y}^{1/2})$$
 with $n - \text{rank}(\Sigma_{\hat{x}_y})$ columns of $V$ spanning $\ker(\Sigma_{\hat{x}_y})$
$$D \leftarrow \text{diag}\begin{Bmatrix} 1 - \beta^{-1}\lambda_i^{-1} & \lambda_i > \beta^{-1} \\ 0 & \lambda_i \leq \beta^{-1} \end{Bmatrix}$$
$$\Sigma_{\hat{x}_u} \leftarrow \Sigma_{\hat{x}_y}^{1/2} V D V^\intercal \Sigma_{\hat{x}_y}^{1/2}$$
**end function**

1) The memory state depends only on the previous memory state and the current observation; that is, $z_t$ is independent of $(z^{t-2}, y^{t-1}, u^{t-1}, x^t)$ given $z_{t-1}$ and $y_t$.
2) The control depends only on the memory state; that is, $u_t$ is independent of $(z^{t-1}, u^{t-1}, x^t, y^t)$ given $z_t$.

A system including a retentive controller satisfies the Bayesian network in Figure 1.

As motivated in Part I for the memoryless case, we are particularly interested in controllers where both the memory state update and the control are linear-Gaussian and time-invariant (LTI), since they are easier to optimize and implement. Linear controllers with limited memory are known not to be optimal for all control problems [8], [9]. The conditions under which such controllers are optimal for our bounded control problem are beyond our current scope.

*Definition 5:* A retentive linear-Gaussian time-invariant (LTI) controller has memory state space that is a vector space

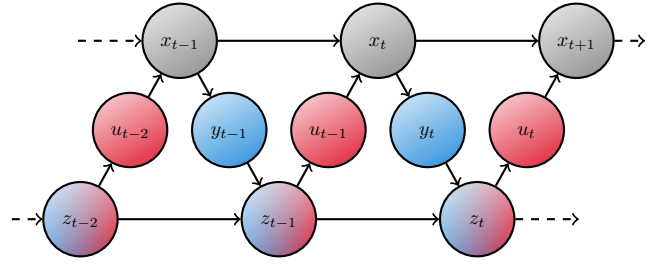

Fig. 1. Bayesian network of retentive control

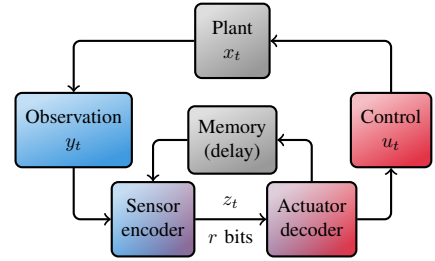

Fig. 2. Block diagram of a closed-loop retentive control system, with a communication channel from the sensor-reader to the actuator-writer

$\mathcal{Z} = \mathbb{R}^d$ and control law of the form

$$z_t = F z_{t-1} + G y_t + \zeta_t; \qquad \zeta_t \sim \mathcal{N}(0, \Sigma_\zeta) \quad (6a)$$
$$u_t = L z_t + \nu_t; \qquad \nu_t \sim \mathcal{N}(0, \Sigma_\nu) \quad (6b)$$

where $F \in \mathbb{R}^{d \times d}$, $G \in \mathbb{R}^{d \times k}$, $\Sigma_\zeta \in \mathbb{S}_+^d$, $L \in \mathbb{R}^{m \times d}$, $\Sigma_\nu \in \mathbb{S}_+^m$, $\zeta_t$ is independent of $(z_{t-1}, y_t)$ and $\nu_t$ is independent of $z_t$.

We are interested in reducing the information complexity of implementing this controller. To measure this complexity, we consider the capacity of a memoryless communication channel from the sensor-reader to the actuator-writer (Figure 2). The encoder and the decoder themselves are memoryless, but the memory component has perfect fidelity, making everything written by the actuator available for the sensor to read in the next step.

We could use $\mathcal{Z} = \{0, 1\}^r$, the set of $r$-bit strings, instead of the vector space $\mathbb{R}^d$, to indicate that the controller can process at most $r$ bits of information per time step

$$\mathbb{I}[z_{t-1}, y_t; z_t, u_t] = \mathbb{I}[z_{t-1}, y_t; z_t] \leq \mathbb{H}[z_t] \leq r \log 2.$$

As in the memoryless case (Part I [1, Section III-B]), the information rate is generally not a tight lower bound on the capacity of a discrete memory, but here again, if the controller is LTI, there exists a perfectly matched memoryless additive Gaussian noise channel. As shown in the Supplementary Material[1] (SM), Appendix I, the capacity of this channel optimally equals the information rate $\mathbb{I}[z_{t-1}, y_t; z_t, u_t]$ and a constraint on the information rate is equivalent to a constraint on the power available for transmission on the channel.

The retentive controller optimization problem is therefore similar to Problem 1, but with the information rate including both the memory and the sensory channels.

---
[1]Available at https://arxiv.org/abs/1606.01947

*Problem 2:* Given a LQG task, the bounded retentive LTI controller optimization problem is

$$\min_{\pi} \quad \mathcal{I}_\pi$$
$$\text{s.t.} \quad \mathcal{J}_\pi \leq c,$$

with $\mathcal{I}_\pi$ as in (5), where

$$\mathcal{I}_t = \mathbb{I}[z_{t-1}, y_t; z_t, u_t], \quad (7)$$

and with $z_t$ and $u_t$ as in (6).

Note that here there is no additional constraint or cost on the precision of $u_t$ given $z_t$, implying that optimally $\Sigma_\nu = 0$.

There is an interesting connection between the retentive information rate $\mathcal{I}_\pi$ and the long-term average of the directed information rate [10], [11], defined by

$$\mathbb{I}[\{y_t\} \to \{z_t\}] = \limsup_{T \to \infty} \frac{1}{T} \mathbb{I}[y^T \to z^T]$$
$$= \limsup_{T \to \infty} \frac{1}{T} \sum_{t=1}^T \mathbb{I}[y^t; z_t | z^{t-1}].$$

By the independence properties of the retentive controller and by the chain rule for information [12], we have

$$\mathbb{I}[z_{t-1}, y_t; z_t, u_t] = \mathbb{I}[z_{t-1}, y_t; z_t]$$
$$= \mathbb{I}[z^{t-1}, y^t; z_t]$$
$$= \mathbb{I}[z^{t-1}; z_t] + \mathbb{I}[y^t; z_t | z^{t-1}].$$

We can thus define the following extension of the concept of directed information.

*Definition 6:* The retentive directed information from the sequence of observations $y^T$ to the sequence of memory states $z^T$ is

$$\mathbb{I}[y^T \twoheadrightarrow z^T] = \sum_{t=1}^T \mathbb{I}[z^{t-1}, y^t; z_t].$$

Since $\mathbb{I}[y^T \twoheadrightarrow z^T] \geq \mathbb{I}[y^T \to z^T]$, the retentive directed information rate is always a tighter lower bound on the capacity of the channel in Figure 2. Despite the apparent similarity to Figure 2 in [11], notice that their encoder and decoder have unlimited memory of $z^t$ and $u^t$. This justifies their use of directed information, regardless of the residual term $\mathbb{I}[z^{t-1}; z_t]$ being infinite in their optimal controller.

Some further properties of the retentive directed information can be found in the SM, Appendix VI.

### B. Reduction to memoryless controllers

We can analyze the bounded retentive control problem (Problem 2) directly using the same tools developed in Part I [1, Section IV-A] for Problem 1. Fortunately, there is no need to repeat that entire treatment, since a simple and insightful reduction will allow us to reuse the results obtained there.

We start by reformulating the problem. The following relaxation and Lemma 1 that shows its equivalence to the original problem allow us to reverse the causality between $u_t$ and $z_t$. We need a new notation for the resulting time-shifted memory state sequence and define for each $t$

$$m_t = z_{t-1}.$$

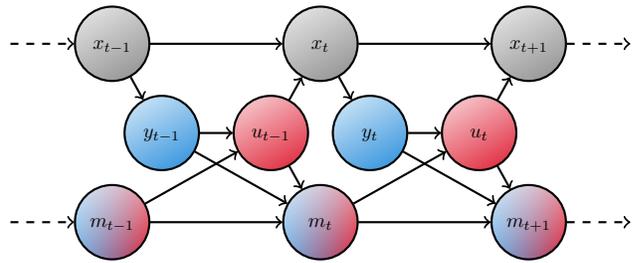

Fig. 3. Bayesian network of relaxed retentive control

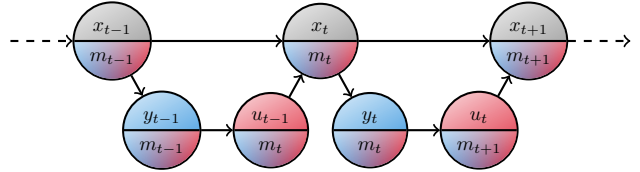

Fig. 4. Bayesian network of relaxed retentive control, redrawn in the form of memoryless control

*Definition 7:* A retentive controller is relaxed if $u_t$ is not required to be independent of $(m_t, y_t)$ given $m_{t+1}$. Thus the relaxed controller satisfies the Bayesian network in Figure 3 and its control law is given by $\pi(u_t, m_{t+1} | m_t, y_t)$.

*Lemma 1:* The relaxed controller optimization problem is equivalent to the original Problem 2.

*Proof:* The following proof does not assume that the controller is linear-Gaussian and holds for the LTI controller as a special case.

Let $\pi$ be a controller satisfying the Bayesian network in Figure 3. We construct a controller $\tilde{\pi}$ with $\tilde{z}_t = (u_t, m_{t+1})$ for each $t$, such that

$$\tilde{\pi}(\tilde{z}_t | \tilde{z}_{t-1}, y_t) = \pi(u_t, m_{t+1} | m_t, y_t)$$
$$\tilde{\pi}(u_t | \tilde{z}_t) = \delta_{\tilde{z}_t = (u_t, \cdot)}.$$

This controller satisfies the Bayesian network in Figure 1 and

$$\mathbb{I}_{\tilde{\pi}}[\tilde{z}_{t-1}, y_t; \tilde{z}_t, u_t] = \mathbb{I}_\pi[(u_{t-1}, m_t), y_t; (u_t, m_{t+1})]$$
$$= \mathbb{I}_\pi[m_t, y_t; u_t, m_{t+1}].$$

Thus the controller $\tilde{\pi}$ is feasible for the unrelaxed Problem 2 and has the same performance as the relaxed controller $\pi$, since it induces a stochastic process with the same distribution and information rate. □

The structure in Figure 3 can now be redrawn as in Figure 4. Comparing this Bayesian network to the one in Part I, Figure 2, we have clearly reduced the bounded retentive control problem to a special case of the bounded memoryless control problem, as stated formally in the following lemma.

*Lemma 2:* The bounded retentive LTI controller optimization problem (Problem 2) for the LQG task $\langle A_x, B_{x;u}, C_{y;x}, \Sigma_\xi, \Sigma_\epsilon, Q_x, R_u \rangle$ is equivalent to the bounded memoryless LTI controller optimization problem (Problem 1) for the LQG task $\langle A, B, C, \Sigma_{\tilde{\xi}}, \Sigma_{\tilde{\epsilon}}, Q, R \rangle$, where

$$A = \begin{bmatrix} A_x & 0 \\ 0 & 0 \end{bmatrix}; \quad B = \begin{bmatrix} B_{x;u} & 0 \\ 0 & I \end{bmatrix}; \quad C = \begin{bmatrix} C_{y;x} & 0 \\ 0 & I \end{bmatrix}$$

$$\Sigma_{\tilde{\xi}} = \begin{bmatrix} \Sigma_\xi & 0 \\ 0 & 0 \end{bmatrix}; \quad \Sigma_{\tilde{\epsilon}} = \begin{bmatrix} \Sigma_\epsilon & 0 \\ 0 & 0 \end{bmatrix}$$

$$Q = \begin{bmatrix} Q_x & 0 \\ 0 & 0 \end{bmatrix}; \quad R = \begin{bmatrix} R_u & 0 \\ 0 & 0 \end{bmatrix}.$$

Here all matrices are extended by $d$ rows and $d$ columns.

*Proof:* Given the retentive control stochastic process $\{x_t, m_t, y_t, u_t\}$, we consider the memoryless control stochastic process $\{\tilde{x}_t, \tilde{y}_t, \tilde{u}_t\}$ with

$$\tilde{x}_t = \begin{bmatrix} x_t \\ m_t \end{bmatrix}; \quad \tilde{y}_t = \begin{bmatrix} y_t \\ m_t \end{bmatrix}; \quad \tilde{u}_t = \begin{bmatrix} u_t \\ m_{t+1} \end{bmatrix}.$$

The dynamics for this process can easily be seen to be given by (1), (2), with $A$, $B$, $C$, $\Sigma_{\tilde{\xi}}$ and $\Sigma_{\tilde{\epsilon}}$ as in the lemma. The cost rate applies only to the $x_t$ and $u_t$ parts

$$\mathcal{J}_t = \tfrac{1}{2}\left( \begin{bmatrix} x_t \\ m_t \end{bmatrix}^\intercal \begin{bmatrix} Q_x & 0 \\ 0 & 0 \end{bmatrix} \begin{bmatrix} x_t \\ m_t \end{bmatrix} + \begin{bmatrix} u_t \\ m_{t+1} \end{bmatrix}^\intercal \begin{bmatrix} R_u & 0 \\ 0 & 0 \end{bmatrix} \begin{bmatrix} u_t \\ m_{t+1} \end{bmatrix} \right).$$

The information rate is

$$\mathcal{I}_t = \mathbb{I}[\tilde{y}_t; \tilde{u}_t] = \mathbb{I}[m_t, y_t; u_t, m_{t+1}],$$

where the left-hand side is taken as in (4) and the right-hand side as in (7), as required. □

### C. Structure of the optimal solution

We can substitute the form of the reduction in Lemma 2 into the optimal solution in Section II-B, to study more explicitly the structure of the optimal solution in the retentive case. The detailed derivations can be found in the SM, Appendix VII.

For the backward process, it is useful to borrow notation from the forward process and denote

$$S = \begin{bmatrix} S_x & S_{x;m} \\ S_{m;x} & S_m \end{bmatrix}$$

$$S_{x|m} = S_x - S_{x;m} S_m^\dagger S_{m;x}$$

$$S_{u|m} = R + B^\intercal S_{x|m} B.$$

Then we can find the feedback gain

$$L = -(R + B^\intercal SB)^\dagger B^\intercal SA$$

$$= \begin{bmatrix} L_{u;x|m} & 0 \\ -S_m^\dagger S_{m;x}(A_x + B_{x;u} L_{u;x|m}) & 0 \end{bmatrix}, \quad (8)$$

with a memory-conditioned form of the classic feedback gain

$$L_{u;x|m} = -S_{u|m}^\dagger B_{x;u}^\intercal S_{x|m} A_x.$$

The memory-conditioned cost reduction matrix is

$$N = L^\intercal (R + B^\intercal SB) L = \begin{bmatrix} N_{x|m} & 0 \\ 0 & 0 \end{bmatrix},$$

with

$$N_{x|m} = A_x^\intercal (S_x - S_{x|m} + S_{x|m} B_{x;u} S_{u|m}^\dagger B_{x;u}^\intercal S_{x|m}) A_x.$$

Thus $\operatorname{rank}(D) \le \operatorname{rank}(N) \le n$, with $D$ the mode activation matrix (see Algorithm 3), implying that at most $n$ modes can be active.

The $d$ rightmost columns in (8) are 0, implying that $\tilde{u}_t$ depends only on the state estimator $\hat{x}_{\tilde{u}_t} = \mathbb{E}[x_t | \tilde{u}_t]$ of $x_t$ and not on an estimator of the memory component $m_t$. Since $\hat{x}_{\tilde{y}_t} = \mathbb{E}[x_t | \tilde{y}_t]$ is a sufficient statistic of $\tilde{y}_t$ for $x_t$, we also have the Markov chain

$$x_t \;\text{---}\; \hat{x}_{\tilde{y}_t} \;\text{---}\; \tilde{y}_t \;\text{---}\; \hat{\tilde{x}}_{\tilde{y}_t} \;\text{---}\; \hat{\tilde{x}}_{\tilde{u}_t} \;\text{---}\; \hat{x}_{\tilde{u}_t} \;\text{---}\; \tilde{u}_t,$$

with

$$\hat{\tilde{x}}_{\tilde{y}_t} = \mathbb{E}[\tilde{x}_t | \tilde{y}_t] = \mathbb{E}\left[ \begin{bmatrix} x_t \\ m_t \end{bmatrix} \bigg| \begin{bmatrix} y_t \\ m_t \end{bmatrix} \right]$$

and similarly for $\hat{\tilde{x}}_{\tilde{u}_t}$. This implies that we need only consider the first component $\hat{x}_{\tilde{y}_t}$ of $\hat{\tilde{x}}_{\tilde{y}_t}$, which is obtained from the observation $\tilde{y}_t$ using

$$K = \Sigma_x C^\intercal \Sigma_{\tilde{y}}^\dagger$$
$$= \begin{bmatrix} K_{x;y|m} & (I - K_{x;y|m} C_{y;x}) \Sigma_{x;m} \Sigma_m^\dagger \end{bmatrix},$$

where

$$K_{x;y|m} = \Sigma_{x|m} C_{y;x}^\intercal \Sigma_{y|m}^\dagger$$

is the Kalman gain that performs optimal inference in the classic LQG task [7].

Crucially, we see that $\hat{x}_{\tilde{y}_t}$ depends on $m_t$ only through

$$\hat{x}_{m_t} = \mathbb{E}[x_t | m_t] = \Sigma_{x;m} \Sigma_m^\dagger m_t.$$

This implies that, for a controller $\pi$, we can design an equivalent controller $\pi'$ whose memory state is the MMSE estimator $m'_t = \hat{x}_{m_t}$. The feedback gain for $\pi'$ is

$$L' = \begin{bmatrix} I & 0 \\ 0 & \Sigma_{x;m} \Sigma_m^\dagger \end{bmatrix} L.$$

Note that, since $m'_t$ is a sufficient statistic of $m_t$ for $x_t$, we have $\Sigma_{x|m'} = \Sigma_{x|m}$ and $K_{x;y|m'} = K_{x;y|m}$. Thus

$$K' = \begin{bmatrix} K_{x;y|m} & I - K_{x;y|m} C_{y;x} \end{bmatrix},$$

with $\Sigma_{x;m'} \Sigma_{m'}^\dagger = \Sigma_{m'} \Sigma_{m'}^\dagger$ in the second component omitted due to its redundancy.

The controllers $\pi$ and $\pi'$ generate the same control $u_t$ and thus incur the same external cost. At the same time, since $m'_t$ is a function of $m_t$, by the data-processing inequality the information rate of $\pi'$ is at most that of $\pi$. Thus any controller can be converted into a MMSE controller without loss of performance, allowing us to consider the MMSE controller canonical. In particular, this proves again that $d = n$ is always sufficient for representing the memory state.

We now diverge from the solution given in Section II-B, which has freedom in its choice of memory representation, and is therefore not guaranteed to be a MMSE controller. Instead, we explicitly constrain the controller to be MMSE, which in return enables us to relax some of the conditions given in Section II-B, which are now not necessary (and indeed do not hold at the optimum), as discussed below.

Constraining the controller to be MMSE imposes the structure

$$\Sigma_{\tilde{x}} = \begin{bmatrix} \Sigma_{x|m} + \Sigma_m & \Sigma_m \\ \Sigma_m & \Sigma_m \end{bmatrix},$$

parameterized by $\Sigma_{x|m}$ and $\Sigma_m$. The reduced number of independent parameters leaves $M$ overparameterized (see SM, Appendix VII) and we can choose, without loss of performance, the structure

$$M = \begin{bmatrix} M_{x|m} + M_m & -M_m \\ -M_m & M_m \end{bmatrix}$$

with

$$M_{x|m} = \beta^{-1} Z$$
$$M_m = \beta^{-1}(C_{y;x}^\mathsf{T} K_{x;y|m}^\mathsf{T} Z K_{x;y|m} C_{y;x} - Z),$$

where $Z = \Sigma_{\hat{x}_{\tilde{y}}|\hat{x}_{\tilde{u}}}^\dagger - \Sigma_{\hat{x}_{\tilde{y}}}^\dagger$ is the signal-to-noise-ratio (SNR) matrix for the channel $\hat{x}_{\tilde{y}_t} \to \hat{x}_{\tilde{u}_t}$. Due to the shrinkage effect of $K_{x;y|m} C_{y;x}$

$$M_m \preceq 0 \preceq M_{x|m} + M_m.$$

The Hessian of the cost-to-go now has the form

$$S = Q + A^\mathsf{T} S A - M$$
$$= \begin{bmatrix} Q_x + A_x^\mathsf{T} S_x A_x - M_{x|m} - M_m & M_m \\ M_m & -M_m \end{bmatrix}$$

and the second-order expansion of the cost-to-go, at the optimum, has the form

$$\tilde{x}_t^\mathsf{T} S \tilde{x}_t = x_t^\mathsf{T}(Q_x + A_x^\mathsf{T} S_x A_x - \beta^{-1} Z) x_t \\ - (m_t - x_t)^\mathsf{T} M_m (m_t - x_t).$$

The first term measures the divergence of the state $x_t$ from 0 and the second the divergence of the controller's estimator $m_t$ from the true state $x_t$, which is the expected form for a MMSE controller. Both terms link the SNR matrix $Z$ to the cost reduction. In this form, $S$ is again positive semidefinite, while now $M$ is generally not.

Finally, when $\beta = \infty$, we can recover the classic LQG results. Similarly to Part I [1, Section IV-B], we can substitute $N_{x|m}$ for $\beta^{-1} Z$, to recover the algebraic Riccati equation

$$S_{x|m} = Q_x + A_x^\mathsf{T} S_x A_x - N_{x|m} \\ = Q_x + A_x^\mathsf{T}(S_{x|m} - S_{x|m} B_{x;u} S_{u|m}^\dagger B_{x;u}^\mathsf{T} S_{x|m}) A_x.$$

## IV. EXAMPLE

As a simple example, consider the double mass-spring-damper system in Figure 5, adapted from [13]. The

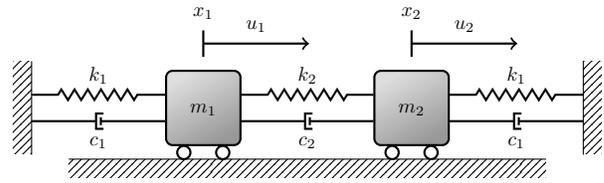

Fig. 5. Double mass-spring-damper system; masses: $m_1 = 5$ kg, $m_2 = \sqrt{15}$ kg; spring constants: $k_1 = 1$ N/m, $k_2 = 0.5$ N/m; damping coefficients: $c_1 = c_2 = 1$ N·sec/m

continuous-time dynamics of this system are given by

$$A = \begin{bmatrix} 0 & 1 & 0 & 0 \\ -\frac{k_1+k_2}{m_1} & -\frac{c_1+c_2}{m_1} & \frac{k_2}{m_1} & \frac{c_2}{m_1} \\ 0 & 0 & 0 & 1 \\ \frac{k_2}{m_2} & \frac{c_2}{m_2} & -\frac{k_1+k_2}{m_2} & -\frac{c_1+c_2}{m_2} \end{bmatrix}$$

$$B = \begin{bmatrix} 0 & 0 \\ \frac{1}{m_1} & 0 \\ 0 & 0 \\ 0 & \frac{1}{m_2} \end{bmatrix} \quad C = \begin{bmatrix} 1 & 0 & 0 & 0 \\ 0 & 0 & 1 & 0 \end{bmatrix},$$

with $m_1 = 5$ kg, $m_2 = \sqrt{15}$ kg, $k_1 = 1$ N/m, $k_2 = 0.5$ N/m and $c_1 = c_2 = 1$ N·sec/m. We discretize the time using the Tustin transformation with sampling frequency 20Hz and consider the isotropic noises and cost rates

$$\Sigma_\xi = I \quad \Sigma_\epsilon = I \quad Q = I \quad R = I.$$

For the memoryless control problem, we initialize a solution with $\Sigma_x = S = 0$. For the retentive control problem, we apply the reduction in Lemma 2 to obtain a reduced plant and then initialize a solution using the classic LQG controller, as described in Section III-C. To the initial solution, we apply the forward-backward iterations of Section II-B, with fixed $\beta$, until convergence to a fixed point, suspected as a global optimum. To improve running time, we employ a reverse-annealing scheme, decreasing $\beta$ gradually over its range and using the fixed point for one value of $\beta$ to initialize the iterations for the next value of $\beta$.

Figures 6 and 7 show, respectively, the resulting cost-log-beta and cost-information curves, demonstrating that even this simple example exhibits interesting phenomenology.

We see that both the memoryless (blue) and the retentive (green) controllers undergo phase transitions as $\beta$ increases. The system is controllable and observable, allowing the retentive controller to undergo 4 phase transitions, until it fully remembers and controls all modes of the system. However, the rank-2 matrices $B$ and $C$ only allow the memoryless controller to undergo 2 phase transitions and reach order $d = 2$.

In the first phase transition, the controllers begin controlling a single mode, in order to reduce the external cost, at the expense of communication resources. This is not depicted in the cost-information plot (Figure 7), since below this critical point the information is 0 and the cost is fixed.

The second and fourth phase transitions involve memory and only occur in the retentive controller. Below these critical point, a hypothetical order-2 retentive controller is worse than the order-1 controller, in terms of the target $\mathcal{F}$, the total external and internal cost-to-go it incurs. At the critical point, the order-2 controller overtakes the order-1 controller, already with a significantly reduced cost rate and a significant information rate (see red dots in Figures 6 and 7). The critical point is where the ratio between these costs is $\beta^{-1}$ (see (12) in Part I [1, Section IV-B]).

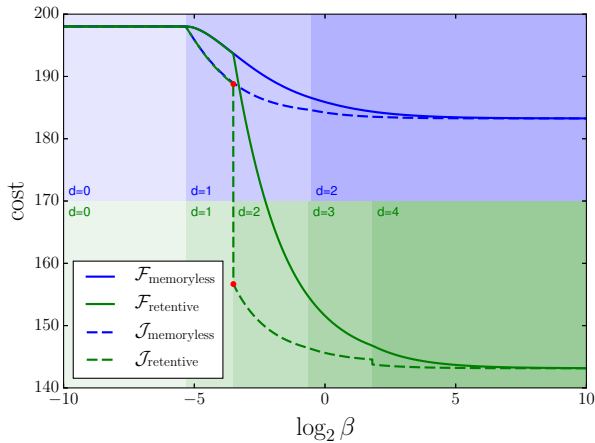 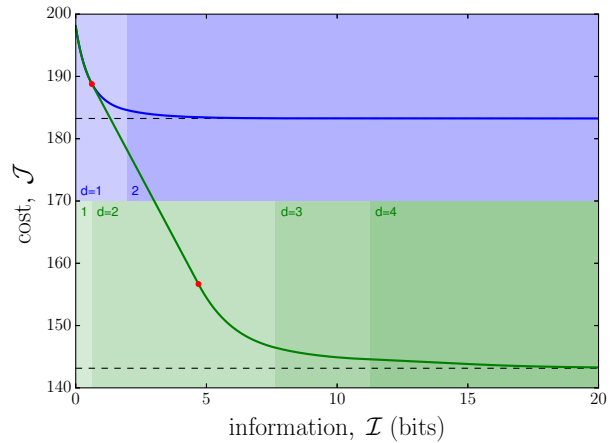

Fig. 6. Cost-log-beta curve for the double mass-spring-damper problem. Memoryless control (blue) generally incurs higher cost than retentive control (green). The Lagrangian $\mathcal{F}$ (solid) is continuous, whereas the external cost $\mathcal{J}$ (dashed) is discontinuous in the retentive case in phase transitions 2 (red dots) and 4. Background shades indicate the controller order $d$, with boundaries at critical points.

Fig. 7. Cost-information curve for the double mass-spring-damper problem. Memoryless control (blue) incurs higher cost than retentive control (green) after phase transition 2 (red dots). The asymptotic costs at $\beta = \infty$ (dashed black) can be approximated with very little information and a reduced order.

The third phase transition is again common to the memoryless and the retentive controllers, although by now the retentive controller has committed to memory much valuable information, reducing the cost much beyond the capabilities of the memoryless controller.

## V. Discussion

In this paper we introduce the problem of optimal LQG control with bounded channel capacity in both the memory and the sensory channels. We show how to reduce this problem to that of bounded memoryless LQG control, study the structure of the resulting solution and illustrate its interesting phenomenology with a simple example.

One aspect of this phenomenology that merits further study is the existence of suboptimal fixed points of the iterative algorithm (Section II-B). For example, around the second critical point in the double mass-spring-damper system (Section IV), both an order-1 controller and a retentive order-2 controller are fixed points. Before the phase transition, one of these solutions is stable, while the other is metastable and suboptimal, and at the phase transition they switch. This resembles well-studied phenomena in statistical physics.

LQG control with constraints on the sensory channel capacity has now been studied in the regime of unlimited memory [11], no memory (Part I of this work [1]) and in this paper, a shared channel capacity for sensing and memory. More generally, the memory and the sensory channels can be separate, with their relative costs ranging from 0 (no memory) to 1 (shared capacity) to $\infty$ (unlimited memory) including any intermediate value. This memory-sensory trade-off has been studied in the context of finite-state systems [4] and further insight can be gained from studying this more general problem in the LQG context.

# Minimum-Information LQG Control Supplementary Material

Roy Fox[†] and Naftali Tishby[†]

## APPENDIX I
## PERFECTLY MATCHED CHANNEL

In this appendix we construct a channel that is perfectly matched to the sequential source code derived in Theorem 1, in Part I of this paper [1, Section III-B]. Recall that in a perfectly matched source-channel pair the optimal source coding and the optimal channel coding can be implemented jointly for single letters, without requiring longer blocks. This allows us to use them in a perception-action cycle, where we cannot accumulate a block of inputs before emitting an output.

The main results of [2], applied to our setting, can be summarized as follows. We wish to find a memoryless channel into which we can input an encoding $w_t = g(\hat{x}_{y_t})$, such that $\hat{x}_{u_t} = h(\hat{w}_t)$ can be decoded from the channel output $\hat{w}_t$. Suppose that we are concerned with the power needed to transmit $w_t$ and thus the input cost is $w_t^\intercal w_t$. Then the source $\hat{x}_{y_t}$ and the channel $w_t \to \hat{w}_t$ are perfectly matched if there exist an encoder and a decoder such that

1) The Kullback-Leibler divergence $\mathbb{D}[f(\hat{w}_t|w_t)\|f(\hat{w}_t)]$ between the conditional and marginal densities of $\hat{w}_t$, as a function of $w_t$, equals $c_1 w_t^\intercal w_t + c_2$, for some constants $c_1 \geq 0$ and $c_2$; and
2) $f(\hat{x}_{u_t}|\hat{x}_{y_t})$ satisfies the conditions in Theorem 1.

To meet these conditions, we can choose the channel, the encoder and the decoder to have

$$w_t = D^{1/2} V^\intercal \Sigma_{\hat{x}_y}^{\dagger/2} \hat{x}_{y_t}$$
$$\hat{w}_t = w_t + v_t; \qquad v_t \sim \mathcal{N}(0, I - D)$$
$$\hat{x}_{u_t} = \Sigma_{\hat{x}_y}^{1/2} V D^{1/2} \hat{w}_t,$$

with $D$ and $V$ as in Theorem 1. Then

$$\Sigma_w = D$$
$$\Sigma_{\hat{w}} = I$$
$$\Sigma_{\hat{x}_u} = \Sigma_{\hat{x}_y}^{1/2} V D V^\intercal \Sigma_{\hat{x}_y}^{1/2} = \Sigma_{\hat{x}_u;\hat{x}_y},$$

and it can be verified that

$$\mathbb{D}[f(\hat{w}_t|w_t)\|f(\hat{w}_t)] = \tfrac{1}{2} w_t^\intercal \Sigma_{\hat{w}}^{-1} w_t + \text{const},$$

as required.

The capacity of the additive Gaussian noise channel with noise covariance $I - D$, under the appropriate expected power constraint, is indeed achieved by a Gaussian input with covariance $D$ and is equal to the information rate in Theorem 1. As shown in [2], this means that constraining the expected power $\Sigma_w$ is equivalent to constraining the information rate $\mathbb{I}[\hat{x}_{y_t}; \hat{x}_{u_t}]$.

Note, however, that the matched channel noise covariance depends on the constraint, through the solution in Theorem 1. Moreover, this result is not applicable when the best channel available to the designer of the controller is not the matched channel above, in which case both the channel and the sequential source coding generally need to be adapted.

## APPENDIX II
## PROOF OF LEMMA 1 OF PART I

In this appendix we restate and prove Lemma 1 of Part I [1, Section IV-A].

*Lemma 1:* Let $x$ and $\hat{x}$ be 0-mean jointly Gaussian random variables. The following properties are equivalent:

1) There exists a random variable $u$, jointly Gaussian with $x$, such that $\hat{x}(u) = \arg\min_{\hat{x}} \mathbb{E}[\|\hat{x} - x\|^2 | u] = \mathbb{E}[x|u]$.
2) $\Sigma_{\hat{x};x} = \Sigma_{\hat{x}}$.
3) $\Sigma_{x|\hat{x}} = \Sigma_x - \Sigma_{\hat{x}}$, where $\Sigma_{x|\hat{x}}$ is the conditional covariance matrix of $x$ given $\hat{x}$, implying $\Sigma_x \succeq \Sigma_{\hat{x}}$.
4) $\hat{x} = \mathbb{E}[x|\hat{x}]$.

Such $\hat{x}$ is called a minimum mean square error (MMSE) estimator (of $u$) for $x$.

*Proof:* ($1 \implies 2$) Assume without loss of generality that $u$ has mean 0. Then

$$\hat{x} = \Sigma_{x;u} \Sigma_u^\dagger u,$$

implying

$$\Sigma_{\hat{x};x} = \Sigma_{x;u} \Sigma_u^\dagger \Sigma_{u;x} = \Sigma_{\hat{x}}.$$

($2 \implies 3$)

$$\Sigma_{x|\hat{x}} = \Sigma_x - \Sigma_{x;\hat{x}} \Sigma_{\hat{x}}^\dagger \Sigma_{\hat{x};x} = \Sigma_x - \Sigma_{\hat{x}}.$$

($3 \implies 4$) Since $x$ and $\hat{x}$ are 0-mean and jointly Gaussian, we can write for some $T$

$$x = T\hat{x} + \xi; \qquad \xi \sim \mathcal{N}(0, \Sigma_{x|\hat{x}}),$$

implying

$$\Sigma_x = T \Sigma_{\hat{x}} T^\intercal + \Sigma_x - \Sigma_{\hat{x}},$$

thus without loss of generality $T = I$.

[†]School of Computer Science and Engineering, The Hebrew University, {royf,tishby}@cs.huji.ac.il
[*]This work was supported by the DARPA MSEE Program, the Gatsby Charitable Foundation, the Israel Science Foundation and the Intel ICRI-CI Institute

($4 \implies 1$) Taking $u = \hat{x}$, we have

$$\arg\min_{\hat{x}'} \mathbb{E}[\|\hat{x}' - x\|^2 | u]$$
$$= \arg\min_{\hat{x}'} (\hat{x}'^\mathsf{T} \hat{x}' - 2\hat{x}'^\mathsf{T} \mathbb{E}[x|u]) + \mathbb{E}[x^\mathsf{T} x | u],$$

which is optimized by $\hat{x}' = \mathbb{E}[x|u]$. $\square$

## APPENDIX III
### PROOF OF LEMMA 2 OF PART I

In this appendix we restate and prove Lemma 2 of Part I [1, Section IV-A].

*Lemma 2:* The bounded memoryless LTI controller optimization problem (Problem 1) is solved by a control law of the form

$$\hat{x}_{y_t} = K y_t \tag{5a}$$
$$\hat{x}_{u_t} = W \hat{x}_{y_t} + \omega_t; \qquad \omega_t \sim \mathcal{N}(0, \Sigma_\omega) \tag{5b}$$
$$u_t = L \hat{x}_{u_t}, \tag{5c}$$

where $W \in \mathbb{R}^{n \times n}$, $\Sigma_\omega \in \mathbb{R}^{n \times n}$, $L \in \mathbb{R}^{\ell \times n}$, $\omega_t$ is independent of $y_t$, $\hat{x}_{u_t}$ is a MMSE estimator for $\hat{x}_{y_t}$ and

$$\mathbb{I}[y_t; u_t] = \mathbb{I}[\hat{x}_{y_t}; \hat{x}_{u_t}]. \tag{6}$$

*Proof:* Consider a LTI controller $\pi$ of the form

$$u_t = H y_t + \eta_t; \qquad \eta_t \sim \mathcal{N}(0, \Sigma_\eta), \tag{III.1}$$

satisfying the Markov network

$$\begin{array}{ccc} x_t \,\text{---}\, & y_t & \text{---}\, u_t \\ & | & | \\ & \hat{x}_{y_t} & \hat{x}_{u_t}. \end{array}$$

We now construct a controller $\pi'$ with control law $u_t'$ based on the estimator $\hat{x}_{u_t}'$ by defining the Markov chain

$$x_t \,\text{---}\, y_t \,\text{---}\, \hat{x}_{y_t} \,\text{---}\, u_t'' \,\text{---}\, \hat{x}_{u_t}' \,\text{---}\, u_t'$$

such that each consecutive pair of variables has the same joint distribution as their unprimed namesakes. Since $\hat{x}_{y_t}$ is a sufficient statistic of $y_t$ for $x_t$, we have the Markov chain $x_t \,\text{---}\, \hat{x}_{y_t} \,\text{---}\, y_t \,\text{---}\, u_t$, implying that $u_t''$ has the same joint distribution with $x_t$ as $u_t$ does. Likewise, $\hat{x}_{u_t}'$ has the same joint distribution with $x_t$ as $\hat{x}_{u_t}$ does. Since $\hat{x}_{u_t}$ is a sufficient statistic of $u_t$ for $x_t$, we have that $u_t'$ also has the same joint distribution with $x_t$ as $u_t$ does.

Thus the controller $\pi'$ induces the same stochastic process $\{x_t, u_t'\}$ and the same external cost. Note that $u_t'$ may not have the same joint distribution with $y_t$ as $u_t$ does and due to the data-processing inequality [3]

$$\mathbb{I}[y_t; u_t] \geq \mathbb{I}[\hat{x}_{y_t}; u_t] = \mathbb{I}[\hat{x}_{y_t}; u_t'']$$
$$\geq \mathbb{I}[\hat{x}_{y_t}; \hat{x}_{u_t}'] \geq \mathbb{I}[y_t; u_t'].$$

Therefore $\pi'$ performs at least as well as $\pi$ and equally well when $\pi$ is optimal, proving (6).

$\hat{x}_{u_t}'$ is a MMSE estimator for $\hat{x}_{y_t}$ since

$$\mathbb{E}[\hat{x}_{y_t} | \hat{x}_{u_t}'] = \mathbb{E}[\mathbb{E}[x_t | y_t] | \hat{x}_{u_t'}]$$
$$= \mathbb{E}[x_t | \hat{x}_{u_t}'] = \hat{x}_{u_t}',$$

where the second equality follows from $x_t \,\text{---}\, y_t \,\text{---}\, \hat{x}_{u_t}'$.

Finally, it may not be clear from the above analysis that $u_t'$ is optimally deterministic in $\hat{x}_{u_t}'$. If $u_t$ has covariance $\Sigma_\nu$ given $\hat{x}_{u_t}'$, the Lagrangian of the optimization problem ((9) in Part I) depends on $\Sigma_\nu$ only through the terms

$$\tfrac{1}{2}(\mathrm{tr}(R \Sigma_\nu) + \mathrm{tr}(S B \Sigma_\nu B^\mathsf{T})).$$

Since $R + B^\mathsf{T} S B \succeq 0$ is positive semidefinite, we can take $\Sigma_\nu = 0$ without loss of performance, recovering the structure (5). Intuitively, the argument is that any noise added to $u_t'$, beyond $\hat{x}_{u_t}'$, is not helpful in compressing $x_t$ and can only increase the external cost without saving any communication cost.

In the other direction, let $u_t$ satisfy the form of Lemma 2. We can rewrite $u_t$ in the form (III.1), with

$$H = LWK$$
$$\Sigma_\eta = L \Sigma_\omega L^\mathsf{T}. \qquad \square$$

## APPENDIX IV
### PROOF OF THEOREM 1 OF PART I

In this appendix we restate and prove Theorem 1 of Part I [1, Section IV-A], which relies on the following Lagrangian developed there.

$$\mathcal{F}_{\Sigma_x, \Sigma_{\hat{x}_u}, L, S; \beta} = \tfrac{1}{2}(\beta^{-1}(\log |\Sigma_{\hat{x}_y}|_\dagger - \log |\Sigma_{\hat{x}_y | \hat{x}_u}|_\dagger) \tag{9}$$
$$+ \mathrm{tr}(Q \Sigma_x) + \mathrm{tr}(RL \Sigma_{\hat{x}_u} L^\mathsf{T})$$
$$+ \mathrm{tr}(S((A+BL)\Sigma_{\hat{x}_u}(A+BL)^\mathsf{T}$$
$$+ A \Sigma_{x|\hat{x}_u} A^\mathsf{T} + \Sigma_\xi - \Sigma_x))).$$

*Theorem 1:* Given $\beta$, the Lagrangian (9) is minimized by a controller satisfying the forward equations

$$\Sigma_x = (A+BL)\Sigma_{\hat{x}_u}(A+BL)^\mathsf{T} \tag{10a}$$
$$\qquad + A \Sigma_{x|\hat{x}_u} A^\mathsf{T} + \Sigma_\xi$$
$$\Sigma_y = C \Sigma_x C^\mathsf{T} + \Sigma_\epsilon \tag{10b}$$
$$K = \Sigma_x C^\mathsf{T} \Sigma_y^\dagger \tag{10c}$$
$$\Sigma_{\hat{x}_y} = K \Sigma_y K^\mathsf{T}, \tag{10d}$$

the backward equations

$$M = \beta^{-1} C^\mathsf{T} K^\mathsf{T} (\Sigma_{\hat{x}_y | \hat{x}_u}^\dagger - \Sigma_{\hat{x}_y}^\dagger) KC \tag{10e}$$
$$S = Q + A^\mathsf{T} S A - M, \tag{10f}$$
$$L = -(R + B^\mathsf{T} S B)^\dagger B^\mathsf{T} S A \tag{10g}$$
$$N = L^\mathsf{T} (R + B^\mathsf{T} S B) L \tag{10h}$$

and the control-based estimator covariance

$$\Sigma_{\hat{x}_u} = \Sigma_{\hat{x}_y}^{1/2} V D V^\mathsf{T} \Sigma_{\hat{x}_y}^{1/2}, \tag{10i}$$

the latter determined by the eigenvalue decomposition (EVD)

$$V \Lambda V^\mathsf{T} = \Sigma_{\hat{x}_y}^{1/2} N \Sigma_{\hat{x}_y}^{1/2} \tag{10j}$$

having $V$ orthogonal with $n - \mathrm{rank}(\Sigma_{\hat{x}_y})$ columns spanning the kernel of $\Sigma_{\hat{x}_y}$ and $\Lambda = \mathrm{diag}\{\lambda_i\}$ and by the active mode coefficient matrix

$$D = \mathrm{diag}\left\{ \begin{array}{ll} 1 - \beta^{-1} \lambda_i^{-1} & \lambda_i > \beta^{-1} \\ 0 & \lambda_i \leq \beta^{-1} \end{array} \right\}. \tag{10k}$$

*Proof:* The minimum of the Lagrangian (9) must satisfy the first-order optimality conditions, i.e. that the gradient with respect to each parameter is 0 at the optimum. We start by differentiating $\mathcal{F}$ by the feedback gain $L$

$$\partial_L \mathcal{F}_{\Sigma_x, \Sigma_{\hat{x}_u}, L, S; \beta} = RL\Sigma_{\hat{x}_u} + B^\mathsf{T} S(A+BL)\Sigma_{\hat{x}_u} = 0,$$

which we rewrite as

$$(R + B^\mathsf{T} SB)L\Sigma_{\hat{x}_u} = -B^\mathsf{T} SA\Sigma_{\hat{x}_u}.$$

As this equation shows, $L$ is underdetermined in the kernel of $\Sigma_{\hat{x}_u}$, since these modes are always 0 in $\hat{x}_{u_t}$ and have no effect on $u_t$. $L$ is also underdetermined in the kernel of $R + B^\mathsf{T} SB$, since these modes have no cost (immediate or future) and can be controlled in any way without affecting the solution's performance. Thus without loss of performance we can take

$$L = -(R + B^\mathsf{T} SB)^\dagger B^\mathsf{T} SA.$$

We substitute this solution back into the Lagrangian, to get

$$\mathcal{F}_{\Sigma_x, \Sigma_{\hat{x}_u}, S; \beta} = \tfrac{1}{2}(\beta^{-1}(\log|\Sigma_{\hat{x}_y}|_\dagger - \log|\Sigma_{\hat{x}_y|\hat{x}_u}|_\dagger) \quad \text{(IV.1)}$$
$$+ \operatorname{tr}(M\Sigma_x) - \operatorname{tr}(N\Sigma_{\hat{x}_u}) + \operatorname{tr}(S\Sigma_\xi)),$$

with

$$M = Q + A^\mathsf{T} SA - S$$
$$N = L^\mathsf{T}(R + B^\mathsf{T} SB)L$$
$$= A^\mathsf{T} SB(R + B^\mathsf{T} SB)^\dagger B^\mathsf{T} SA.$$

The problem of optimizing over $\Sigma_{\hat{x}_u}$ given the other parameters can now be written, up to constants, as the semidefinite program (SDP)

$$\max_{\Sigma_{\hat{x}_u}} \quad \log|\Sigma_{\hat{x}_y} - \Sigma_{\hat{x}_u}|_\dagger + \beta \operatorname{tr}(N\Sigma_{\hat{x}_u})$$
$$\text{s.t.} \quad 0 \preceq \Sigma_{\hat{x}_u} \preceq \Sigma_{\hat{x}_y}.$$

By Lemma V.1 in Appendix V, the optimum is achieved when $\Sigma_{\hat{x}_u}$ satisfies (10i)–(10k).

Finally, with $P = \Sigma_{\hat{x}_y} \Sigma_{\hat{x}_y}^\dagger$ the projection onto the support of $\hat{x}_{y_t}$ and since the range of $\Sigma_{\hat{x}_u}$ is contained in that subspace, we have

$$\partial_{(\Sigma_x)_{i,j}}(\log|\Sigma_{\hat{x}_y}|_\dagger - \log|\Sigma_{\hat{x}_y|\hat{x}_u}|_\dagger)$$
$$= -\partial_{(\Sigma_x)_{i,j}} \log|P - \Sigma_{\hat{x}_u} \Sigma_{\hat{x}_y}^\dagger|_\dagger$$
$$= -\partial_{(\Sigma_x)_{i,j}} \log|I - \Sigma_{\hat{x}_u}(P\Sigma_{\hat{x}_y}P)^\dagger|$$
$$= \operatorname{tr}((I - \Sigma_{\hat{x}_u}\Sigma_{\hat{x}_y}^\dagger)^{-1}\Sigma_{\hat{x}_u}\partial_{(\Sigma_x)_{i,j}}(P\Sigma_{\hat{x}_y}P)^\dagger).$$

The purpose of introducing $P$ is to notice that even if the range of $\Sigma_{\hat{x}_y}$ is increased, this has no effect on the Lagrangian, because these modes are orthogonal to the range of $\Sigma_{\hat{x}_u}$. This allows us to treat $P$ as constant, so that the range of $P\Sigma_{\hat{x}_y}P$ is constant in a neighborhood of the solution, and the derivative of the pseudoinverse is simplified in this case to

$$\partial_{(\Sigma_x)_{i,j}}(P\Sigma_{\hat{x}_y}P)^\dagger = -\Sigma_{\hat{x}_y}^\dagger (\partial_{(\Sigma_x)_{i,j}} \Sigma_{\hat{x}_y}) \Sigma_{\hat{x}_y}^\dagger$$
$$= -\Sigma_{\hat{x}_y}^\dagger KCJ_{i,j}C^\mathsf{T} K^\mathsf{T} \Sigma_{\hat{x}_y}^\dagger,$$

with $J_{i,j}$ the matrix with 1 in position $(i,j)$ and 0 elsewhere. This yields

$$\partial_{\Sigma_x} \mathcal{F}_{\Sigma_x, \Sigma_{\hat{x}_u}, S; \beta}$$
$$= \tfrac{1}{2}(M - \beta^{-1}C^\mathsf{T} K^\mathsf{T} \Sigma_{\hat{x}_y}^\dagger (I - \Sigma_{\hat{x}_u} \Sigma_{\hat{x}_y}^\dagger)^{-1} \Sigma_{\hat{x}_u} \Sigma_{\hat{x}_y}^\dagger KC)$$
$$= \tfrac{1}{2}(M - \beta^{-1}C^\mathsf{T} K^\mathsf{T} \Sigma_{\hat{x}_y}^\dagger ((I - \Sigma_{\hat{x}_u} \Sigma_{\hat{x}_y}^\dagger)^{-1} - I)KC)$$
$$= \tfrac{1}{2}(M - \beta^{-1}C^\mathsf{T} K^\mathsf{T} (\Sigma_{\hat{x}_y|\hat{x}_u}^\dagger - \Sigma_{\hat{x}_y}^\dagger)KC) = 0,$$

implying (10e). □

## APPENDIX V
## SEMIDEFINITE PROGRAM SOLUTION

In this appendix we state and prove the following solution to our SDP problem.

*Lemma V.1:* The semidefinite program

$$\max_{X \in \mathbb{S}_+^n} \quad \log|M_1 - X|_\dagger + \operatorname{tr}(M_2 X)$$
$$\text{s.t.} \quad X \preceq M_1,$$

with $M_1, M_2 \succeq 0$, is optimized by

$$X = M_1^{1/2} VDV^\mathsf{T} M_1^{1/2},$$

with the eigenvalue decomposition (EVD)

$$V\Lambda V^\mathsf{T} = M_1^{1/2} M_2 M_1^{1/2},$$

such that $V$ is orthogonal with $n - \operatorname{rank}(M_1)$ columns spanning the kernel of $M_1$ and $\Lambda = \operatorname{diag}\{\lambda_i\}$ and with

$$D = \operatorname{diag} \left\{ \begin{array}{ll} 1 - \lambda_i^{-1} & \lambda_i > 1 \\ 0 & \lambda_i \leq 1 \end{array} \right\}.$$

*Proof:* Let the EVD of $M_1$ be

$$U\Psi U^\mathsf{T} = M_1,$$

with $U$ orthogonal and $\Psi$ diagonal, having

$$\Psi = \begin{bmatrix} \Psi_+ & 0 \\ 0 & 0_{(n-m) \times (n-m)} \end{bmatrix},$$

with $m = \operatorname{rank}(M_1)$. Let

$$\Psi^\ddagger = \Psi^\dagger + I - \Psi^\dagger \Psi = \begin{bmatrix} \Psi_+^{-1} & 0 \\ 0 & I \end{bmatrix}.$$

By changing the variable to

$$Y = \Psi^{\ddagger/2} U^\mathsf{T} X U \Psi^{\ddagger/2},$$

the constraint of the SDP becomes

$$Y \preceq I_{m,n} = \begin{bmatrix} I_{m \times m} & 0 \\ 0 & 0_{(n-m) \times (n-m)} \end{bmatrix}.$$

$Y$ must therefore be 0 outside the upper-left $m \times m$ block, and the SDP is equivalent, up to constants, to

$$\max_{Y \in \mathbb{S}_+^n} \quad \log|I_{m,n} - Y|_\dagger + \operatorname{tr}(\Psi^{1/2} U^\mathsf{T} M_2 U \Psi^{1/2} Y)$$
$$\text{s.t.} \quad Y \preceq I_{m,n}.$$

Let the EVD of the linear coefficient be

$$\bar{V}\Lambda \bar{V}^\mathsf{T} = \Psi^{1/2} U^\mathsf{T} M_2 U \Psi^{1/2},$$

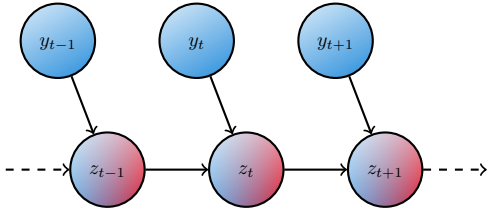

Fig. VI.1. Bayesian network of online inference from a sequence of independent observations

with
$$\bar{V} = \begin{bmatrix} \bar{V}_+ & 0 \\ 0 & I_{(n-m)\times(n-m)} \end{bmatrix}$$
orthogonal and preserving the kernel of $\Psi$ and $\Lambda = \mathrm{diag}\{\lambda_i\}$. We can again change the variable to
$$D = \bar{V}^\mathsf{T} Y \bar{V},$$
to get
$$\max_{D \in \mathbb{S}^n_+} \log|I_{m,n} - D|_\dagger + \mathrm{tr}(\Lambda D)$$
$$\text{s.t.} \quad D \preceq I_{m,n},$$
which can easily be solved using Hadamard's inequality [3], to find
$$D = \mathrm{diag}\left\{\begin{array}{ll} 1 - \lambda_i^{-1} & \lambda_i > 1 \\ 0 & \lambda_i \leq 1 \end{array}\right\}.$$

Finally, the lemma follows by unmaking the variable changes and taking
$$V = U\bar{V}. \qquad \square$$

## APPENDIX VI
## PROPERTIES OF THE RETENTIVE DIRECTED INFORMATION

In this appendix we show how the retentive directed information (Definition 6 of Part II [4, Section III-A]) relates to the multi-information of Bayesian networks [5].

Consider the Bayesian network in Figure VI.1, which describes the process of online inference from a sequence of independent observations. The multi-information of this network, for horizon $T$, is equal to the retentive directed information
$$\mathbb{I}[y^T, z^T] = \mathbb{E}\left[\log \frac{f(y^T, z^T)}{\prod_{t=1}^T f(y_t)f(z_t)}\right]$$
$$= \sum_{t=1}^T \mathbb{E}\left[\log \frac{f(z_t|z^{t-1}, y^t)}{f(z_t)}\right] = \mathbb{I}[y^T \twoheadrightarrow z^T].$$

An important property of the directed information is that the mutual information between two sequences can be decomposed into the sum of directed information in both directions [6]
$$\mathbb{I}[x^T; z^T] = \mathbb{I}[x^T \to z^T] + \mathbb{I}[z^T \to x^T].$$
Interestingly, retentive directed information extends this property to the retentive control process (Figure 1 in Part II). This process can be thought of as consisting of four phases: observation, inference, control and state transition. Its multi-information can accordingly be decomposed [7] into the sum
$$\mathbb{I}[x^T, y^T, z^T, u^T] = \mathbb{I}[x^T \twoheadrightarrow y^T] + \mathbb{I}[y^T \twoheadrightarrow z^T]$$
$$+ \mathbb{I}[z^T \twoheadrightarrow u^T] + \mathbb{I}[u^T \twoheadrightarrow x^T].$$

## APPENDIX VII
## STRUCTURE OF THE OPTIMAL RETENTIVE CONTROLLER

In this appendix we derive the structure of the optimal retentive controller summarized in Part II [4, Section III-C].

For the structured feedback gain $L$ we find using the Schur complement that
$$(R + B^\mathsf{T} S B)^\dagger = \begin{bmatrix} R_u + B_{x;u}^\mathsf{T} S_x B_{x;u} & B_{x;u}^\mathsf{T} S_{x;m} \\ S_{m;x} B_{x;u} & S_m \end{bmatrix}^\dagger$$
$$= \begin{bmatrix} S_{u|m}^\dagger & -S_{u|m}^\dagger B_{x;u}^\mathsf{T} S_{x;m} S_m^\dagger \\ -S_m^\dagger S_{m;x} B_{x;u} S_{u|m}^\dagger & S_{m|u}^\dagger \end{bmatrix},$$
with
$$S_{m|u}^\dagger = S_m^\dagger + S_m^\dagger S_{m;x} B_{x;u} S_{u|m}^\dagger B_{x;u}^\mathsf{T} S_{x;m} S_m^\dagger,$$
and so
$$L = -(R + B^\mathsf{T} SB)^\dagger B^\mathsf{T} SA$$
$$= -(R + B^\mathsf{T} SB)^\dagger \begin{bmatrix} B_{x;u}^\mathsf{T} S_x A_x & 0 \\ S_{m;x} A_x & 0 \end{bmatrix}$$
$$= -\begin{bmatrix} S_{u|m}^\dagger B_{x;u}^\mathsf{T} S_{x|m} A_x & 0 \\ S_m^\dagger S_{m;x}(I - B_{x;u} S_{u|m}^\dagger B_{x;u}^\mathsf{T} S_{x|m}) A_x & 0 \end{bmatrix}$$
$$= \begin{bmatrix} L_{u;x|m} & 0 \\ -S_m^\dagger S_{m;x}(A_x + B_{x;u} L_{u;x|m}) & 0 \end{bmatrix},$$
with
$$L_{u;x|m} = -S_{u|m}^\dagger B_{x;u}^\mathsf{T} S_{x|m} A_x.$$
We also have
$$N = L^\mathsf{T}(R + B^\mathsf{T} SB)L$$
$$= A^\mathsf{T} SB(R + B^\mathsf{T} SB)^\dagger B^\mathsf{T} SA = \begin{bmatrix} N_{x|m} & 0 \\ 0 & 0 \end{bmatrix}$$
$$N_{x|m} = \begin{bmatrix} B_{x;u}^\mathsf{T} S_x A_x \\ S_{m;x} A_x \end{bmatrix}^\mathsf{T} L \begin{bmatrix} I \\ 0 \end{bmatrix}$$
$$= A_x^\mathsf{T}(S_x - S_{x|m} + S_{x|m} B_{x;u} S_{u|m}^\dagger B_{x;u}^\mathsf{T} S_{x|m}) A_x.$$

Dually, for the structured Kalman gain $K$ we find that
$$\Sigma_{\tilde{y}}^\dagger = \begin{bmatrix} \Sigma_y & \Sigma_{y;m} \\ \Sigma_{m;y} & \Sigma_m \end{bmatrix}^\dagger$$
$$= \begin{bmatrix} \Sigma_{y|m}^\dagger & -\Sigma_{y|m}^\dagger \Sigma_{y;m} \Sigma_m^\dagger \\ -\Sigma_m^\dagger \Sigma_{m;y} \Sigma_{y|m}^\dagger & \Sigma_m^\dagger + \Sigma_m^\dagger \Sigma_{m;y} \Sigma_{y|m}^\dagger \Sigma_{y;m} \Sigma_m^\dagger \end{bmatrix},$$
and so
$$K = \Sigma_x C^\mathsf{T} \Sigma_{\tilde{y}}^\dagger$$
$$= \begin{bmatrix} \Sigma_x C_{y;x}^\mathsf{T} & \Sigma_{x;m} \end{bmatrix} \begin{bmatrix} \Sigma_y & \Sigma_{y;m} \\ \Sigma_{m;y} & \Sigma_m \end{bmatrix}^\dagger$$
$$= \begin{bmatrix} K_{x;y|m} & (I - K_{x;y|m} C_{y;x}) \Sigma_{x;m} \Sigma_m^\dagger \end{bmatrix},$$

with
$$K_{x;y|m} = \Sigma_{x|m} C_{y;x}^\intercal \Sigma_{y|m}^\dagger.$$

Now constraining the controller to be MMSE, we have the structure
$$\Sigma_{\tilde{x}} = \begin{bmatrix} \Sigma_{x|m} + \Sigma_m & \Sigma_m \\ \Sigma_m & \Sigma_m \end{bmatrix}$$
$$K = \begin{bmatrix} K_{x;y|m} & I - K_{x;y|m} C_{y;x} \end{bmatrix},$$

which we employ in differentiating $\mathcal{F}$ (IV.1), to get

$$\partial_{\Sigma_{x|m}} \mathcal{F}_{\Sigma_{x|m}, \Sigma_m, \Sigma_{\hat{x}_{\tilde{u}}}, S; \beta} = \begin{bmatrix} I \\ 0 \end{bmatrix}^\intercal \partial_{\Sigma_{\tilde{x}}} \mathcal{F}_{\Sigma_{\tilde{x}}, \Sigma_{\hat{x}_{\tilde{u}}}, S; \beta} \begin{bmatrix} I \\ 0 \end{bmatrix}$$
$$= \tfrac{1}{2} \begin{bmatrix} I \\ 0 \end{bmatrix}^\intercal (M - \beta^{-1} C^\intercal K^\intercal Z K C) \begin{bmatrix} I \\ 0 \end{bmatrix}$$
$$= \tfrac{1}{2} \left( \begin{bmatrix} I \\ 0 \end{bmatrix}^\intercal M \begin{bmatrix} I \\ 0 \end{bmatrix} - \beta^{-1} C_{y;x}^\intercal K_{x;y|m}^\intercal Z K_{x;y|m} C_{y;x} \right) = 0$$
$$\partial_{\Sigma_m} \mathcal{F}_{\Sigma_{x|m}, \Sigma_m, \Sigma_{\hat{x}_{\tilde{u}}}, S; \beta} = \begin{bmatrix} I \\ I \end{bmatrix}^\intercal \partial_{\Sigma_{\tilde{x}}} \mathcal{F}_{\Sigma_{\tilde{x}}, \Sigma_{\hat{x}_{\tilde{u}}}, S; \beta} \begin{bmatrix} I \\ I \end{bmatrix}$$
$$= \tfrac{1}{2} \begin{bmatrix} I \\ I \end{bmatrix}^\intercal (M - \beta^{-1} C^\intercal K^\intercal Z K C) \begin{bmatrix} I \\ I \end{bmatrix}$$
$$= \tfrac{1}{2} \left( \begin{bmatrix} I \\ I \end{bmatrix}^\intercal M \begin{bmatrix} I \\ I \end{bmatrix} - \beta^{-1} Z \right) = 0,$$

with
$$Z = \Sigma_{\hat{x}_{\tilde{y}}|\hat{x}_{\tilde{u}}}^\dagger - \Sigma_{\hat{x}_{\tilde{y}}}^\dagger.$$

This leaves $M$ overparameterized and we can choose to give it the structure
$$M = \begin{bmatrix} M_{x|m} + M_m & -M_m \\ -M_m & M_m \end{bmatrix}$$

with
$$M_{x|m} = \beta^{-1} Z$$
$$M_m = \beta^{-1} (C_{y;x}^\intercal K_{x;y|m}^\intercal Z K_{x;y|m} C_{y;x} - Z).$$